\documentclass[twocolumn,epj,nopacs]{svjour}

\usepackage{amsfonts}
\usepackage{amsmath}
\usepackage{amssymb}
\usepackage{graphicx}
\usepackage{mathrsfs}
\usepackage{hhline,color}
\usepackage{pifont,ulem}
\usepackage{lmodern}
\usepackage{epsfig}
\usepackage{enumitem} 
\usepackage{ulem}
\usepackage{hyperref}
\usepackage[utf8]{inputenc}

\def\beq{\begin{equation}}
\def\eeq{\end{equation}}
\def\beqa{\begin{eqnarray}}
\def\eeqa{\end{eqnarray}}
\def\ban{\begin{eqnarray*}}
\def\ean{\end{eqnarray*}}
\def\bi{\begin{itemize}}
\def\ei{\end{itemize}}

\newcommand{\Z}{\mathbb{Z}}

\begin{document}
\title{Isentropic thermodynamics and scalar mesons properties near the QCD 
critical end point
}

\author{Pedro Costa 
\thanks {email: pcosta@uc.pt}
}                     
%
%
\institute{CFisUC, Department of Physics,
University of Coimbra, P-3004 - 516  Coimbra, Portugal}
\date{Received: date / Revised version: date}

\authorrunning{P. Costa} 
\titlerunning{Isentropic thermodynamics and scalar mesons properties near the QCD 
critical end point}

\abstract{
We investigate the QCD phase diagram and the location of the critical end point 
(CEP) in the SU(2) Polyakov$-$Nambu$-$Jona-Lasinio model with entanglement 
interaction giving special attention to the $\pi$ and $\sigma$-mesons properties, 
namely the decay widths $\sigma\rightarrow\pi\pi$, for several conditions 
around the CEP:
we focus on the possible $\sigma\rightarrow\pi\pi$ decay along the 
isentropic trajectories close to the CEP since the hydrodynamical expansion of 
a heavy-ion collision fireball nearly follows trajectories of constant entropy.
It is expected that the type of transition the dense medium goes through as it 
expands after the thermalization determines the behavior of this decay.
It is shown that no pions are produced from the sigma decay in the chirally 
symmetric phase if the isentropic lines approach the first order line from 
chemical potentials above it. 
Near the CEP or above the $\sigma\rightarrow\pi\pi$ decay is possible with a 
high decay width.
%
} 
\maketitle

\section{Introduction}
\label{intro}

The possible existence of the critical end point (CEP) and its implications to 
the investigation of the QCD phase diagram is a very timely topic that has drawn 
the attention of the physics community. 

From the experimental point of view, the location of the CEP is one major goal 
of several heavy ion collisions (HIC) programs. 
At RHIC, the Beam Energy Scan (BES-I) program, ongoing since 2010, is looking 
for the experimental signatures of the first-order phase transition and 
the CEP by colliding Au ions at several energies \cite{Abelev:2009bw}. 
Also the STAR Collaboration presented their measurements on the moments of 
net-charge multiplicity distributions, which can provide relevant information on 
the freeze-out conditions, in order to clarify the existence of the CEP, but no 
definitive conclusions were possible and future measurements with high 
statistics data will be needed \cite{Adamczyk:2014fia}.
With the upcoming BES-II program, it is expected that, if the CEP exists at a 
baryonic chemical potential below 400 MeV, it can provide data on 
fluctuation and flow observables which should yield quantitative evidence for 
the presence of the CEP.

Meanwhile, the NA49 program at CERN SPS has also investigated the CEP's 
location in nuclear collisions at 158A GeV \cite{Anticic:2009pe}: the analysis 
for $\pi^+ \pi^-$ pairs with an invariant mass very close to the 
two-pion threshold has been performed in \cite{Anticic:2009pe}. 
This sector is important because it may reveal critical fluctuations of the 
sigma component in a hadronic medium, even if the $\sigma$- meson has no 
well-defined vacuum state. In spite of a sizable effect of $\pi^+\pi^-$ pair 
fluctuations with critical characteristics found in Si + Si collisions
at 158A GeV, this effect could not be directly related to the presence of 
the CEP. Now, the NA61/SHINE program is devoted to the search for 
the CEP and to investigate the properties of the onset of deconfinement 
in light and heavy ion collisions \cite{Gazdzicki:2011fx}.
So far, no definitive results were found about the existence of 
the CEP. 

In the next years, FAIR facility at GSI and the Nuclotron-based Ion Collider
Facility at JINR (NICA) will extend the CEP's search to even higher $\mu_B$ 
and definitive conclusions concerning its existence and location are expected 
(for a review on the experimental search of the CEP see \cite{Akiba:2015jwa}).

It is known that the location of the CEP is affected by several conditions 
like the isospin or strangeness content of the medium \cite{Costa:2013zca}, the 
presence of an external magnetic field \cite{Costa:2013zca,Costa:2015bza} or the 
role of the vector interaction in the medium 
\cite{Costa:2015bza,Fukushima:2008wg}. 
The determination of the CEP's location will set stringiest constraint on 
effective models. 

Probes like diphoton and dipion productions are important 
tools for the search of the CEP. 
In this work we will focus on the dipion production. However, concerning 
the diphoton production, it is important to point out that the enhancement of
photon pair production rate at threshold should also belong to the set of
observable effects for the investigation of chiral symmetry restoration in 
ultra-relativisitc HIC experiments. 
Indeed, in \cite{Rehberg:1997xe} it was shown that the process 
$\bar{q}q\rightarrow\gamma\gamma$, which occurs due to the formation of mesonic
resonances, leads to an enhancement of photon pairs with invariant mass equal to 
the thermal pion mass.
On the other hand, the photon pair production by pion annihilation 
($2\pi\rightarrow2\gamma$) at the chiral phase transition was investigated
in \cite{Volkov:1997dx} where the following results were found: a strong
enhancement of the cross section for the pion annihilation process when compared 
with the vacuum case; the calculation of the photon pair production rate as
function of the invariant mass showed a strong enhancement and narrowing of the
sigma meson resonance at threshold. Both results are directly related to
the chiral symmetry restoration.
However, the $\pi^0\rightarrow\gamma\gamma$ decay gives a strong background
contribution making difficult the observation of these effects at high energy
collisions.
One strength of NICA facility, namely the Baryonic Matter at Nuclotron (BM@N)
experiment, concerns the diphoton production: the electromagnetic calorimeters 
that will be used have large acceptance and high resolution allowing to
investigate the invariant mass spectra of $\gamma\gamma$ pairs in the wide range
at different energies and transverse momenta of pairs \cite{nica_BM_at_N}. 

The measurement of the $\sigma\rightarrow\gamma\gamma$ decay is also relevant
because of the small final state interactions \cite{Fukushima:2002mp}. 
Finally, measuring the $\sigma\rightarrow 2\pi^0\rightarrow 4\gamma$ 
\cite{Hatsuda:1999kd} can avoid the possible background from the $\rho-$ meson 
inherent to the $\pi^+\pi^-$ measurement. 

In the present work we investigate the $\pi$ and $\sigma$-mesons masses and the 
decay width of $\sigma\rightarrow\pi\pi$ for several conditions around the CEP,
namely along the isentropic trajectories close to the CEP because it is likely
that the system expands nearly isentropically after the thermalization.

\section{Model and formalism}
\label{sec:Model}

In this work we consider the two-flavor Polyakov$-$Nambu$-$ Jona-Lasinio (PNJL) 
model which Lagrangian  is
\begin{eqnarray}
    {\cal L} \;&=&\; {\bar q}(i\gamma^\mu D^{\mu} - {\hat m_0}) q
    \;+\; G_S\,\Big[({\bar q} q)^2 + ({\bar q}i\gamma_5\vec{\tau} q)^2 \Big] \nonumber\\
    &+& \mathcal{U}\left(\Phi[A],\bar\Phi[A];T\right).
\label{LPNJL}
\end{eqnarray}
Here, $q = (u,d)^T$ represents a quark field with 2-flavors, 
${\hat m}_0= {\rm diag}_f (m_u,m_d)$ is the corresponding (current) mass 
matrix, with $m_u = m_d = m_0$ (to keep the isospin symmetry) and 
$\vec{\tau}$ is a Pauli matrix which acts in flavor space.

The quarks are coupled to the gauge sector {\it via} the covariant derivative, 
$D^{\mu}=\partial^\mu-iA^\mu$,
and the order parameter of $\Z_3$ symmetric/broken phase transition in pure gauge is
the Polyakov loop $\Phi$ (see details in \cite{Ratti:2005jh,Hansen:2006ee}). 

The pure gauge sector is described by the effective potential 
$\mathcal{U}\left(\Phi,\bar\Phi;T\right)$ chosen in order to reproduce the 
results obtained in lattice calculations. From the several possibilities, we
use \cite{Ratti:2005jh},
\begin{eqnarray}
    \frac{\mathcal{U}\left(\Phi,\bar\Phi;T\right)}{T^4}
   &=&-\frac{a\left(T\right)}{2}\bar\Phi \Phi + b(T)\nonumber\\ 
   &\times&\mbox{ln}[1-6\bar\Phi \Phi+4(\bar\Phi^3+ \Phi^3)-3(\bar\Phi\Phi)^2],
    \label{Ueff}
\end{eqnarray}
where
\begin{equation}
    a\left(T\right)=a_0+a_1\left(\frac{T_0}{T}\right)+a_2\left(\frac{T_0}{T}
  \right)^2\,\mbox{and }\,b(T)=b_3\left(\frac{T_0}{T}\right)^3.
\end{equation}
The parameters for this effective potential are $a_0 = 3.51$, $a_1=-2.47$, 
$a_2=15.2$, and $b_3= -1.75$.
$T_0$, the critical temperature for the deconfinement phase transition within 
pure gauge, can be fixed to $T_0=270$ MeV according to lattice findings.
However, this value of $T_0$ leads to a difference between the chiral and 
deconfinement transition temperatures (indicating a week entanglement between 
both transitions), and also a larger value of the 
deconfinement temperature at a zero chemical potential than the value 
$T_c=173\pm 8$ MeV given by full LQCD data \cite{Karsch:2000kv}. 

The extended version of the model where an effective four-quark vertex depending 
on the Polyakov loop is introduced, the entanglement-PNJL (EPNJL) model 
\cite{Sakai:2010rp}, allows the reduction of the difference between the 
transitions temperatures.  
We implement an explicit dependence of $G_S$ on $\Phi$ and $\bar{\Phi}$ assuming 
the following form:
\begin{equation}
G_S(\Phi) = G_S^0\left[1\,-\,\alpha_1\,\Phi \bar\Phi\,-\,\alpha_2\,( \Phi^3\,+
\,\bar\Phi^3) \right],
\label{gphi}
\end{equation}
which respects chiral, $P$, $C$ and the extended $\Z_3$ symmetries 
(the values $\alpha_1=\alpha_2=0.2$ were fixed in \cite{Sakai:2010rp}).

The Polyakov potential $\mathcal{U}$ may depend on 
$\mu$ ($\mu=\mu_q=(\mu_u+\mu_d)/2$) as a consequence of the backreaction of the 
fermion sector to the gluon sector. 
This dependence of $\mathcal{U}$ in $\mu$ can be introduced by a $T_0(\mu)$ 
which was estimated from renormalization-group arguments \cite{Herbst:2010rf}:
\begin{equation}
T_0 (\mu,N_f) = T_{\tau} e^{-\frac{1}{\alpha_0 b(\mu)}}
\label{T_O}
\end{equation}
with $b(\mu)=(11N_c-2N_f)/(6\pi)-16N_f\mu^2/(\pi T^2_{\tau})$, $\alpha_0=0.304$ 
and $T_{\tau} = 1.770$ (GeV). 
The value of $T_0$ also depends on $N_f$ and in our case ($N_f=2$) this leads to
$T_0(\mu=0,N_f)=208$ MeV, with an uncertainty not small than $\pm 30$ MeV
\cite{Herbst:2010rf}. 

From the Lagrangian (\ref{LPNJL}) it is straightforward to obtain the PNJL grand 
potential density in the mean-field approximation, the constituent 
quark mass and the quark condensate, $\left\langle \bar{q}q\right\rangle$
(for a detailed description see \cite{Ratti:2005jh,Hansen:2006ee}).

As a regularization procedure we use a sharp cutoff, $\Lambda$, in 
three-momentum space for all integrals. 
In the numerical calculations our parameters are: $m_0 = 6$ MeV, 
$\Lambda = 590$ MeV, and $G_S^0\Lambda^2 = 2.435$ which give 
$M_q^{vac} = 400$ MeV, $m_{\pi} = 140.2$ MeV, $f_{\pi}=92.6$ MeV 
\cite{Buballa:2003qv}. We will use $T_0(\mu=0,N_f=2)=179$ MeV, that is the 
lower limit of the estimation of $T_0(\mu=0,N_f=2)$, which allows the transition 
temperature for the desconfinement to be close to the LQCD value. 
$T_0(\mu,N_f=2)$ will follow Eq. (\ref{T_O}).

\section{Phase diagram, the location of the CEP and isentropic trajectories}
\label{sec:phase_diagram}

\begin{figure}
\centering
\resizebox{0.5\textwidth}{!}{%
  \includegraphics{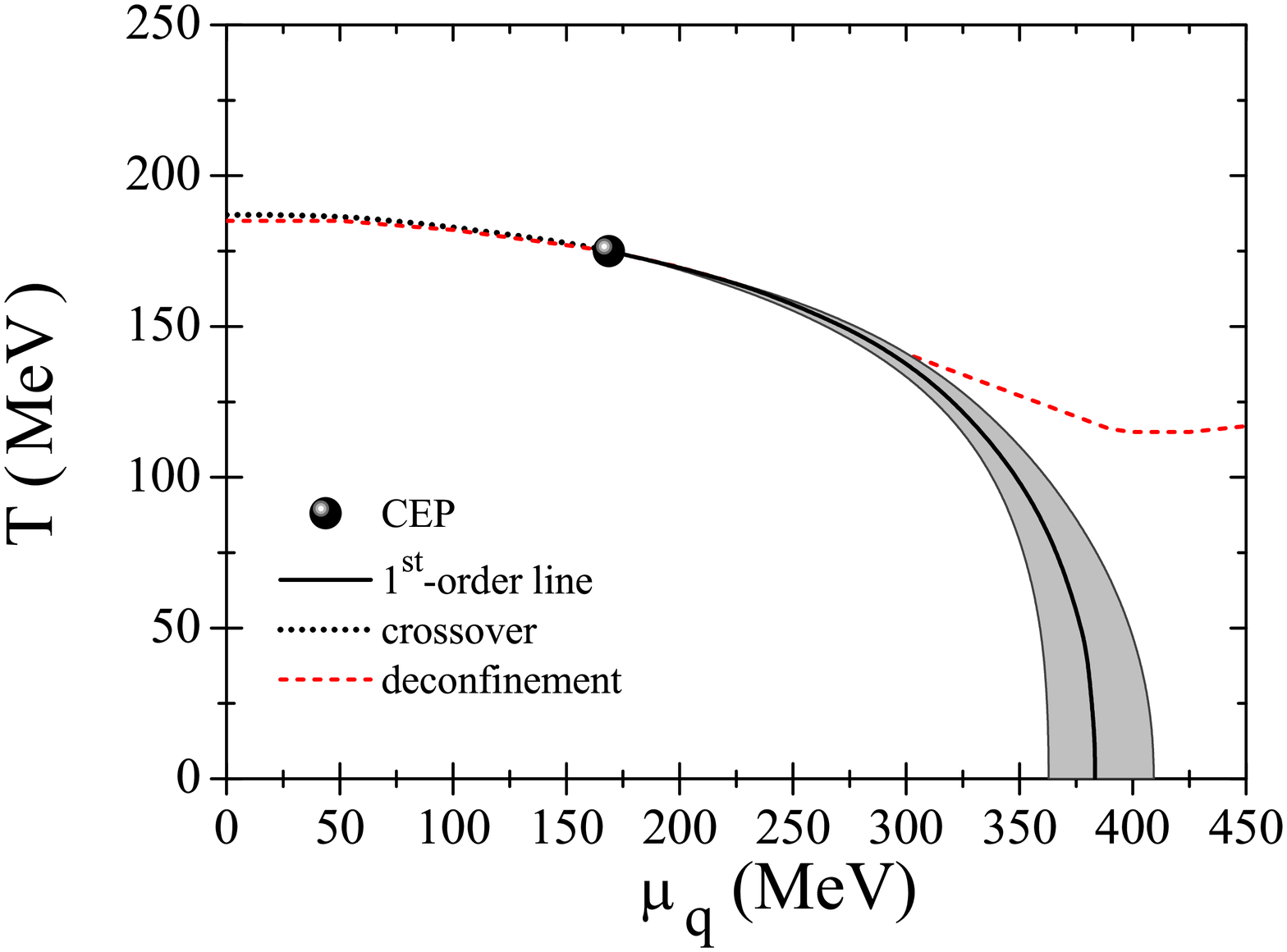}
}
\caption{
QCD phase diagram in the $T-\mu_q$ plane. 
The red dashed line corresponds to the deconfinement transition.
The full black line is the first order chiral phase transition and the gray
region is the spinodal region.
}
\label{fig:1}       
\end{figure}

The phase diagram is plotted in Fig. \ref{fig:1} and is determined by the
grand canonical potential dependence on the order parameters 
$\left\langle \bar{q}q\right\rangle$, $\Phi$ and $\bar{\Phi}$ as a  
function of temperature and chemical potential.
The deconfinement transition is defined  as $\partial^2\Phi/\partial T^2=0$ 
(dotted lines)\footnote{We could also represent 
$\partial^2{\bar\Phi}/\partial T^2=0$ but we will restrict ourselves to
$\partial^2\Phi/\partial T^2=0$.} while the crossover line is defined as 
$\partial^2\left\langle \bar{q}q\right\rangle/\partial T^2=0$ (dashed lines).

At $\mu=0$ the deconfinement transition and the chiral crossover almost 
coincide at $T=185$ MeV. 
At finite $T$ and $\mu$ the deconfinement transition and the chiral crossover 
stay almost coincident (see red and black dashed lines, respectively) until 
the CEP is reached at ($T^{CEP}=175$ MeV; $\mu_q^{CEP}=169$ MeV). 
When $T=0$, with the chosen parametrization, the first order phase transition 
occurs at $\mu_q^{crit} = 383$ MeV.
As $\mu_q^{crit} < M_q^{vac}$ this allows for the existence of quark droplets 
(states in mechanical equilibrium with the vacuum state, $\rho_B=0$, at $P=0$) 
\cite{Buballa:2003qv}.

The isentropic lines (contours of constant entropy per ba\-ry\-on $s/\rho_B$) 
contain important information on the adiabatic evolution of the system. 
This has important consequences in HIC because the expansion of 
the QGP, which is accepted to be an hydrodynamic expansion of an ideal 
fluid, will nearly follow trajectories of constant entropy. 
Indeed, the fast (local) thermalization time and the good agreement of the 
data at RHIC with ideal relativistic hydrodynamic models (assuming a fluid 
evolution with zero viscosity) have been presented as evidences that the matter 
formed at RHIC is a strongly interacting QGP \cite{d'Enterria:2006su}.
Due to its relevance, we investigate the isentropic lines crossing the 
chiral phase transition around the CEP in both the crossover and first order 
transitions. 

\begin{figure}[t]
\centering{
\resizebox{0.5\textwidth}{!}{%
\includegraphics{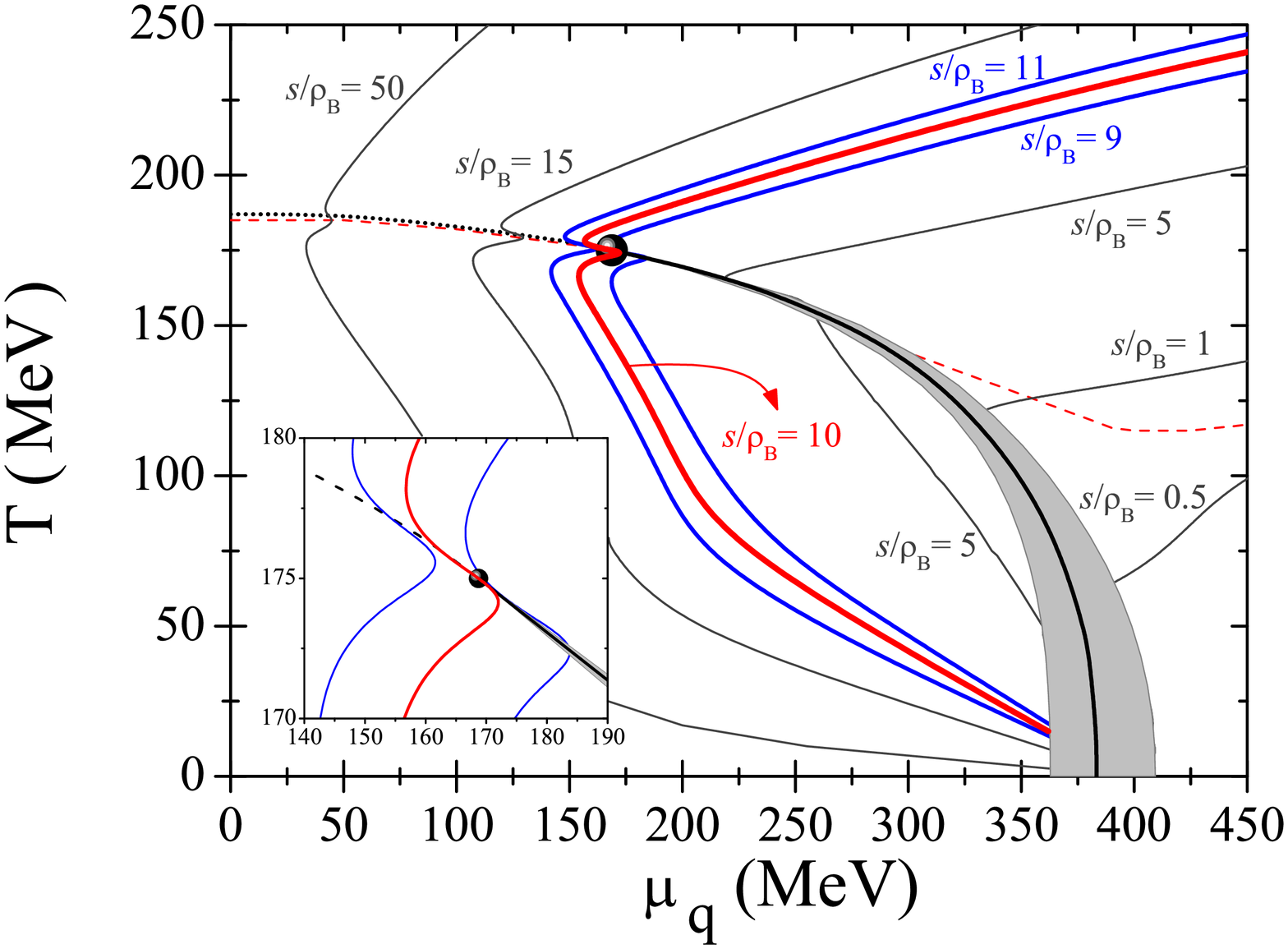}
}
}
\caption{
Isentropic trajectories in the $T-\mu_q$ plane.
}
\label{fig:2}       
\end{figure}

The results for the isentropic lines in the ($T-\mu_q$) plane are shown in 
Fig. \ref{fig:2}. 
We first analyse the behavior of the isentropic lines in the limit 
$T\rightarrow0$.  
As already pointed out, our model allows for the existence of quark droplets, 
and, in addition, simple thermodynamic behavior in the limit $T\rightarrow 0$ 
are verified. 
Indeed, in this limit $s\rightarrow 0$ according to the third law of 
thermodynamics and the condition $s/\rho_B = const.$ is satisfied 
\cite{Costa:2009ae}.
Near the first order region, the isentropic lines with $s/\rho_B \lesssim 4$ 
come from the region of partial restored chiral symmetry and reach directly 
the phase transition region going then along with it as $T$ decreases 
until it reaches $T = 0$. 
The isentropic lines with $4\lesssim s/\rho_B \lesssim 10$ intersect the first
order line and go over it for a while. Then, they leave the first order region 
through the chirally broken phase and they will reach the first order region
again for higher (lower) values of $\mu$ $(T)$. 
The trajectory with $s/\rho_B = 10$ arrives to the first order line very 
close the CEP and shows a pronounced kink behavior, a ``focusing'' effect already
found in \cite{Nonaka:2004pg}. The same happens to the case of $s/\rho_B = 11$ 
(the CEP acts as an attractor of isentropic trajectories \cite{Nonaka:2004pg}).

In the crossover region the isentropic trajectories have slight kink behaviors 
when crossing the transition and they reach the first order region from lower 
values of $\mu_q$. For isentropic lines $s/\rho_B \approx 50$ the kink almost 
disappears and the trajectories have a qualitatively similar 
behavior to the one obtained in lattice calculations \cite{Ejiri:2005uv}. 

All trajectories directly terminate at the same point of the horizontal axes at 
$T=0$: as the temperature decreases the first order phase transition occurs, 
the latent heat increases and the formation of the mixed phase is 
thermodynamically favored.

\section{$\pi$ and $\sigma$-mesons properties around the CEP}
\label{sec:Mesons}

We can obtain additional information about the phase diagram by calculating the 
masses of the pion and the sigma mesons, $m_{\pi}$ and $m_{\sigma}$, as 
functions of $T$ and $\mu_q$. 
These masses, are obtained by using the standard mesonic polarization functions 
$\Pi_{\pi}$ and $\Pi_{\sigma}$ (see {\cite{Hatsuda:1994pi} for details). 

At $T= \mu_q= 0$ the pion is a bound state, but, as the temperature 
increases it will dissociate in a $\bar{q}q$ pair (when $M_{\pi}> 2M_q$) 
at the Mott temperature. The polarization operator acquires an imaginary part
and the resonance $m_{\pi}$ has an associated decay width.
The sigma is always a resonance and dissociates in a 
$\bar{q}q$ for all temperatures\footnote{We make the zero width approximation and 
only take the real part of the polarization operators \cite{Hatsuda:1987kg}.}.

The behavior of the masses of the chiral partners ($\pi, \sigma$) at 
($T\ne 0, \mu_q= 0$) and at ($T= 0, \mu_q\ne0$) are qualitatively similar and 
well known from the literature: they both converge at a certain value of the 
temperature (chemical potential). This is known as the {\it effective} restoration 
of chiral symmetry and can also be seen by the merging of the $\pi$ and $\sigma$ 
spectral functions \cite{Hansen:2006ee}. 
For $T\ne 0$ and $\mu_q=0$ the degeneracy of the chiral partners occurs in a 
range of temperatures where the mesons are no longer bound states.
Since the pion dissociates in $\bar{q}q$ pair, in Fig. \ref{fig:3} we represent 
the respective ``Mott line'' (red dot-dashed line). Above $\mu_q\approx401$ MeV
 the ``Mott line'' occurs inside the first order region.
Due to its relevance for the $\sigma\rightarrow\pi\pi$ process we also plot in 
Fig. \ref{fig:3} the threshold of this decay defined by 
$m_\sigma (T,\mu)=2m_\pi (T,\mu)$ (blue line).

\begin{figure}
\centering
\resizebox{0.5\textwidth}{!}{%
  \includegraphics{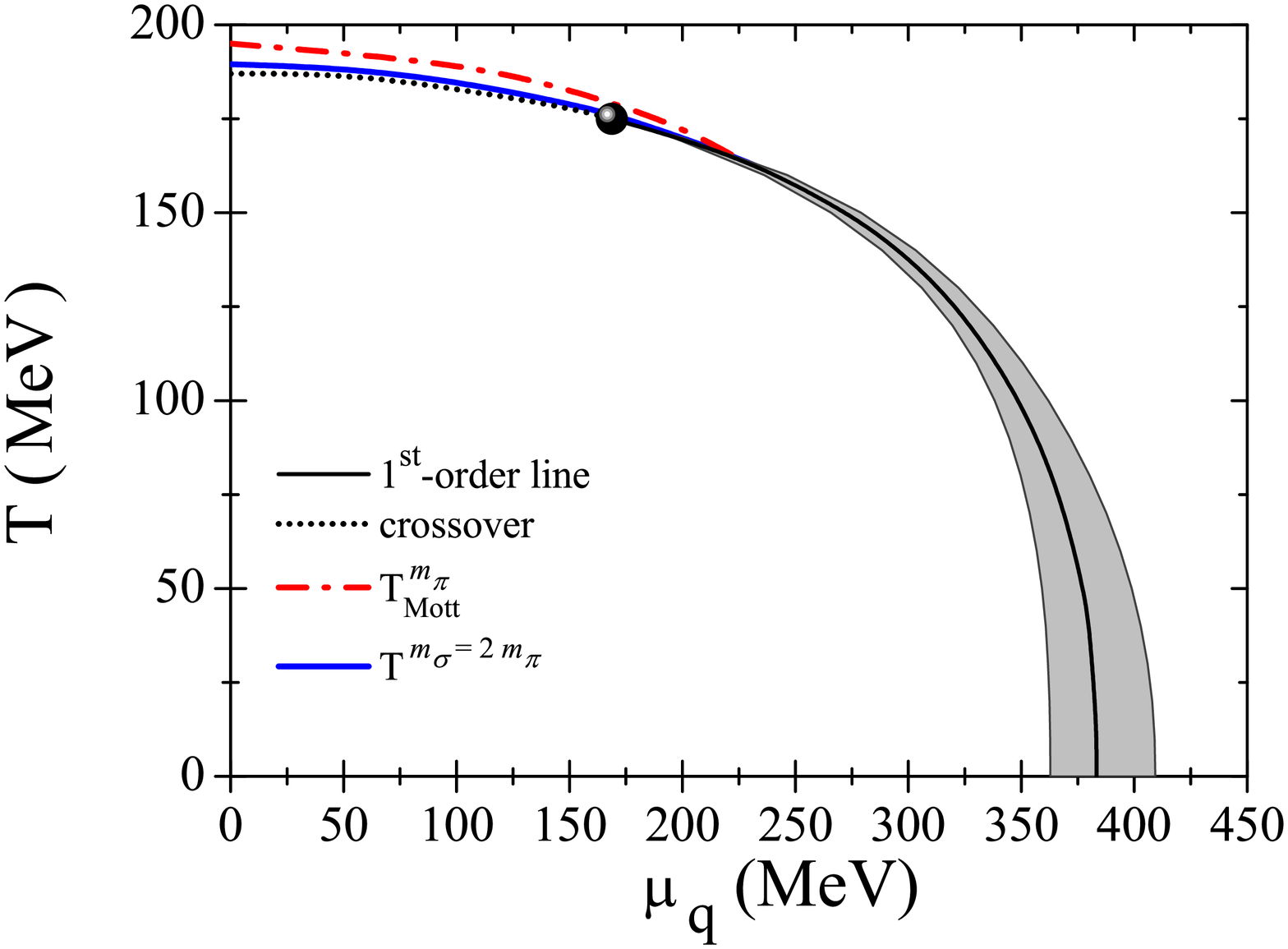}
}
\caption{The chiral phase transition, the line where $m_{\sigma} = 2 m_{\pi}$
(blue line), and the Mott line for the pion (red dot-dashed line) in the 
$T-\mu_q$ plane.
}
\label{fig:3}       
\end{figure}

One interesting aspect is that, around the CEP, the $\sigma$-meson can have an 
abnormally small mass. This means that it is expected that some peculiar 
experimental signatures can be observed through its spectral changes.
In fact, in hot and dense media the $\sigma$-meson can decay through different 
processes like $\sigma\rightarrow\pi\pi$, $\sigma\rightarrow\gamma\gamma$, etc.
We will focus on the $\sigma\rightarrow\pi\pi$ decay but we will not consider 
the Bose-Einstein statistics for the final state pions because, as pointed out in
\cite{Zhuang:2000uc}, this effect at high temperatures is gradually washed out 
as $\mu_q$ increases.

Near the chiral transition temperature, $m_{\sigma}$ is significantly 
reduced while $m_{\pi}$ increases making impossible the $\sigma$-meson decay 
into two pions and, therefore, the width coming from this process vanishes being
the threshold point of the decay defined by $m_\sigma (T,\mu)=2m_\pi (T,\mu)$. 
Indeed, in the chiral limit the threshold
$m_\sigma (T,\mu)=2m_\pi (T,\mu)$ must coincide with the restoration of
chiral symmetry because at that point $M_q=m_0$ and $m_\sigma (T,\mu)=m_\pi (T,\mu)$.
Outside the chiral limit, $M_q$ goes asymptotically to $m_0$ and at the transition
temperature $M_q>>m_0$: both phenomena do not coincide, but they occur near 
each other.

Once we assume the $\sigma$-meson as a quark-antiquark pair, the decay width for 
the $\sigma\rightarrow\pi\pi$ process is:
\begin{eqnarray} 
\Gamma_{\sigma\rightarrow\pi\pi} = \frac{3}{2} \frac{g^2_{\sigma\pi\pi}}
{16\,m_{\sigma}}\sqrt{1 - \frac{4\,m_{\pi}^2}{m_{\sigma}^2}},
\label{GamPiPi}
\end{eqnarray}
where $g_{\sigma\pi\pi}(T, \mu) = 2g_\sigma g^2_\pi A_{\sigma\pi\pi}(T, \mu)$ 
is the coupling stren\-gth.
$g_\sigma$ and $g_\pi$ are coupling constants for the $\sigma$ and 
$\pi$-mesons respectively, and $A_{\sigma\pi\pi}$ is the amplitude of the 
triangle vertex for the decay $\sigma\rightarrow\pi\pi$ (see details in 
\cite{Zhuang:2000uc,Friesen:2011ma,Hufner:1994vd}).
The  constraint $m_\sigma (T,\mu)\leq2m_\pi (T,\mu)$ comes from the factor 
$\sqrt{1 - 4m_{\pi}^2/m_{\sigma}^2}$ in Eq. (\ref{GamPiPi}). When 
$m_\sigma (T,\mu)>2m_\pi (T,\mu)$ the values $g_{\sigma\pi\pi}$ and 
$\Gamma_{\sigma\rightarrow\pi\pi}$ will go to zero.

In Fig. \ref{fig:4} we present the behavior of $\Gamma_{\sigma\rightarrow\pi\pi}$ 
and 
$g_{\sigma\pi\pi}$ as functions of the temperature for 3 cases:\\
Case I $-$ $\mu_q=0$;\\
Case II $-$ $\mu_q=\mu_q^{CEP}$ and $\mu_q=\mu_q^{CEP}\pm 30$ MeV;\\
Case III $-$ $s/\rho_B=9$ (the isentropic line reaches the first order phase 
transition from the chirally symmetric phase);
$s/\rho_B=10$ (the isentropic line passes very close the CEP;
$s/\rho_B=11$ (the isentropic line goes through the crossover and intersects 
the first order line from below).

\begin{figure*}
\centering{
\resizebox{1\textwidth}{!}{%
\includegraphics{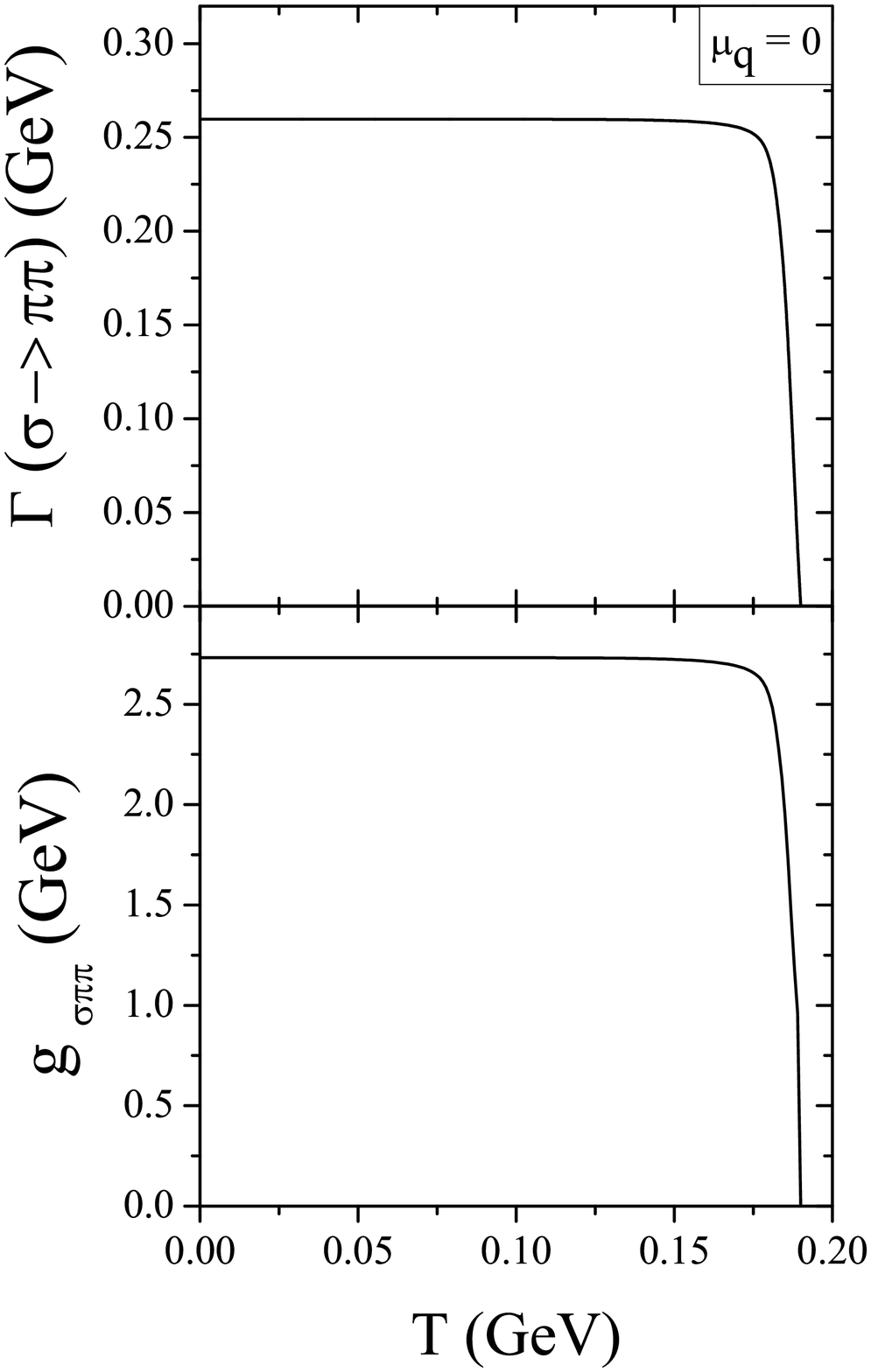}
\includegraphics{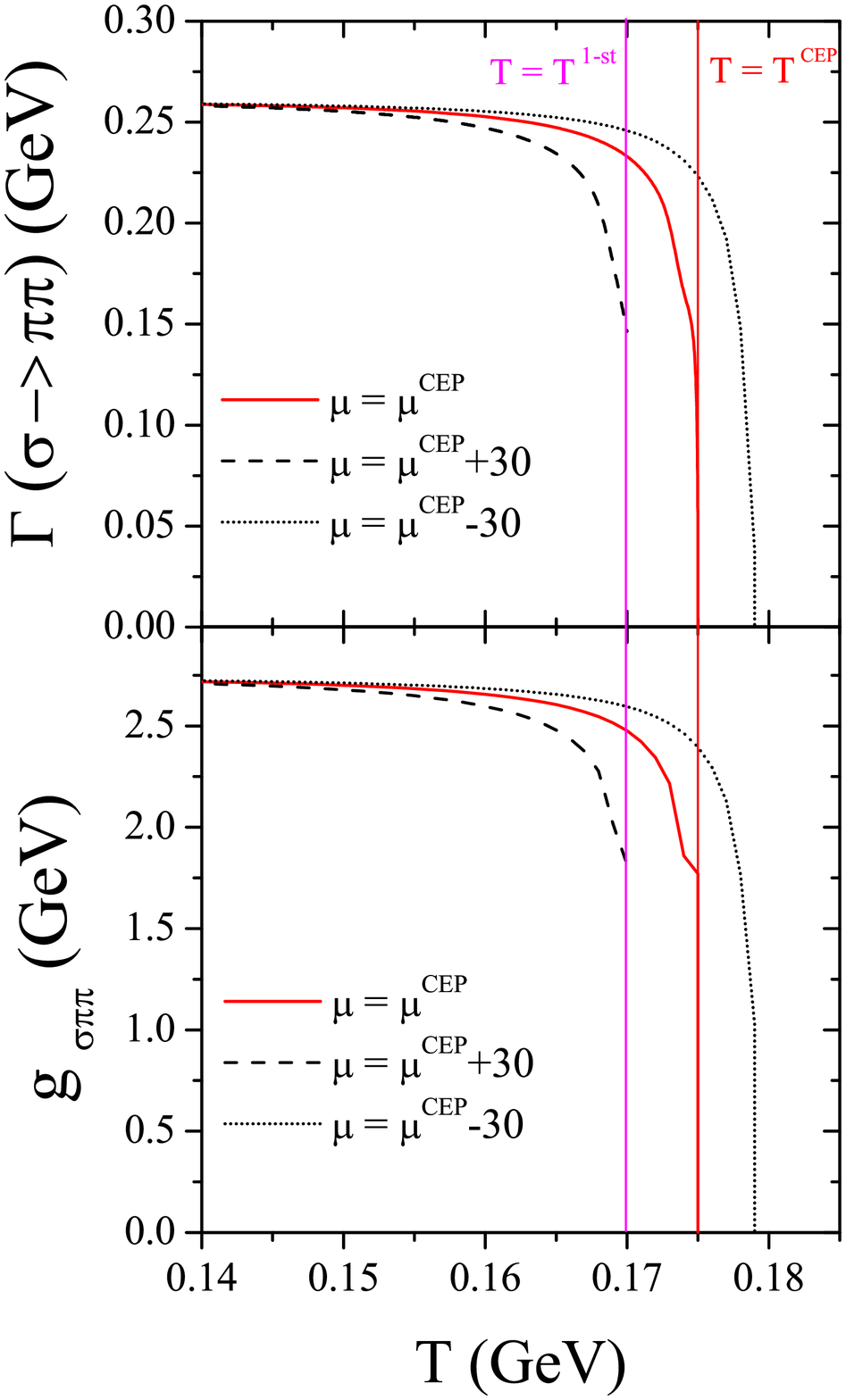}
\includegraphics{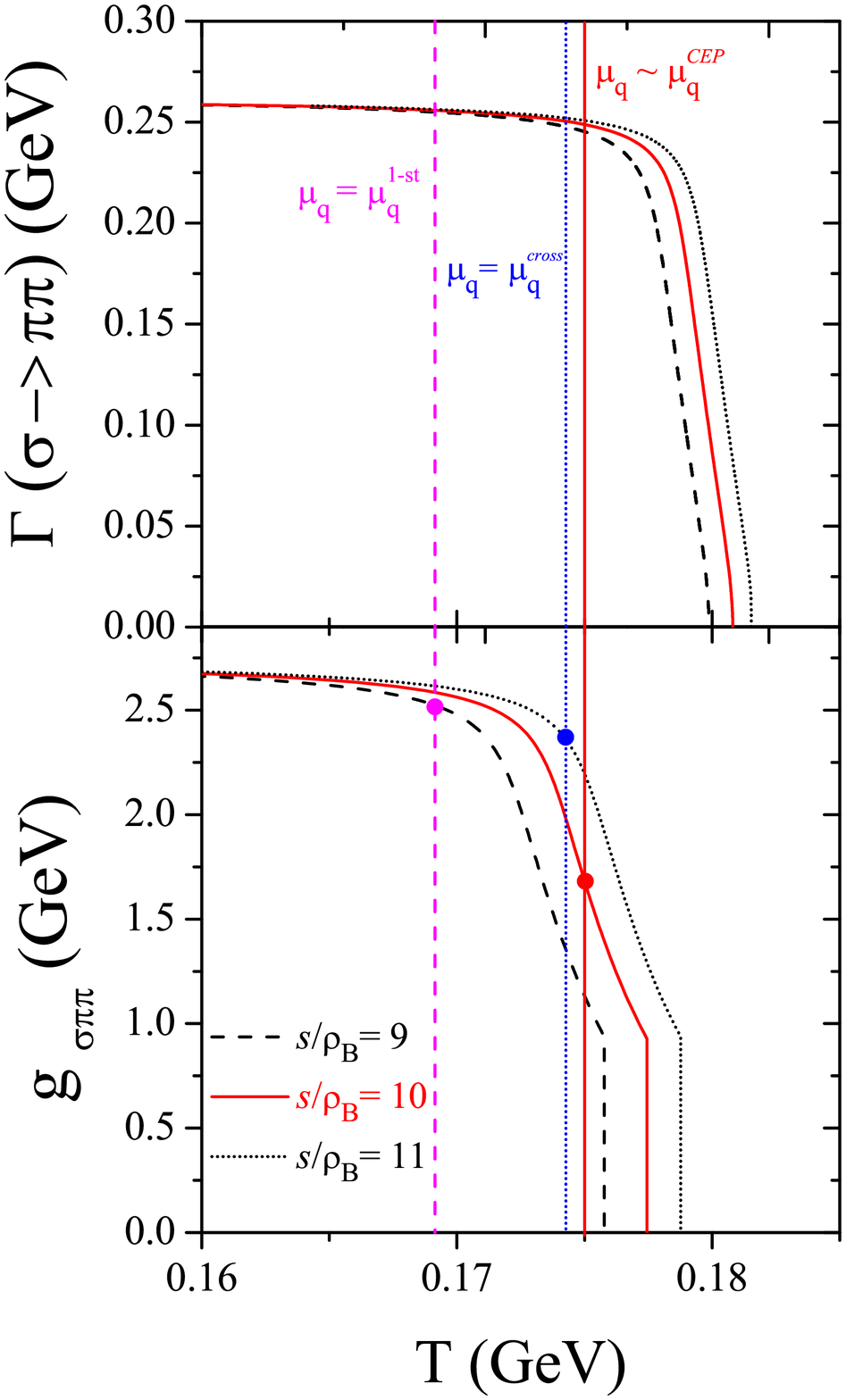}
}
}
\caption{
$\Gamma_{\sigma\rightarrow\pi\pi}$ (upper panels) and $g_{\sigma\pi\pi}$
(lower panels) for the three cases presented: Case I ($\mu_q=0$), left panels;
Case II ($\mu_q=\mu_q^{CEP},\,\mu_q^{CEP}\pm30$), middle panels; Case III
($s/\rho_B=9,10,11$), right panels.
}
\label{fig:4}       
\end{figure*}

In Case I (Fig. \ref{fig:4}, left panel), at $T=0$, we have $m_{\sigma} = 803.8$
MeV which gives $g_{\sigma\pi\pi} = 2.73$ GeV, slightly above the experimental 
value extracted from the J/$\psi$ decays given by the BES collaboration:  
$g_{\sigma\pi\pi}=2.0^{+0.3}_{-0.9}$ GeV~\cite{Wu:2001vz}. 
When the temperature increases $g_{\sigma\pi\pi}$ drops to zero near the point  
$m_{\sigma} = 2m_{\pi}$. 
For the $\sigma\rightarrow\pi\pi$ decay, we obtain 
$\Gamma_{\sigma\rightarrow\pi\pi} = 260$ MeV, within the certainty intervals of 
the experimental results from the BES collaboration 
$\Gamma_{\sigma\rightarrow\pi\pi} = 282^{+77}_{-50}$ MeV 
(with $m_{\sigma} = 390^{+60}_{-36}$ MeV), or the  E791 Collaboration with
$\Gamma_{\sigma\rightarrow\pi\pi} = 324^{+40}_{-42}\pm 21$ MeV (with 
$m_{\sigma} = 478^{+24}_{-23}\pm 17$ MeV) \cite{Aitala:2000xu}. 
As noticed in \cite{Zhuang:2000uc}, the choice of parameters can make shifts of 
magnitudes in 
$g_{\sigma\pi\pi}$ and $\Gamma_{\sigma\rightarrow\pi\pi}$, mainly due to the 
value of $m_{\sigma}$, but does not change their shapes, especially the behavior 
around the threshold temperature or the chiral transition.

In Case II, around the CEP, (Fig. \ref{fig:4}, middle panel) the result is very 
similar to what was found in \cite{Zhuang:2000uc,Friesen:2011ma}: as the 
chemical potential increases close to the CEP, the threshold  temperature the 
for $\sigma$ decay decreases. 
In the first-order transition region ($\mu_q =\mu_q^{CEP}+30$ MeV) 
$\left\langle\bar{q}q \right\rangle$, $m_{\pi}$ and $m_{\sigma}$ are 
discontinuous: $\left\langle\bar{q}q \right\rangle$ and $m_{\sigma}$ jump down 
and $m_{\pi}$ jumps up and, consequently, the mass difference between $m_{\pi}$ 
and $m_{\sigma}$ jumps from below 2$m_{\pi}$ to above 2$m_{\pi}$. 
The $\sigma$ decay threshold coincides with the first-order transition line and 
$g_{\sigma\pi\pi}$ and $\Gamma_{\sigma\rightarrow\pi\pi}$ go to zero.
At the CEP ($\mu_q =\mu_q^{CEP}$) the limiting threshold for the $\sigma$ decay
approximately coincides with the temperature of the CEP, $T^{CEP}$ (see red line
for $g_{\sigma\pi\pi}$ and $\Gamma_{\sigma\rightarrow\pi\pi}$).
In the crossover region ($\mu_q =\mu_q^{CEP}-30$ MeV), $g_{\sigma\pi\pi}$ and 
$\Gamma_{\sigma\rightarrow\pi\pi}$ go to zero slightly above the temperature 
where $\partial^2\left\langle \bar{q}q\right\rangle/\partial T^2=0$. 
This is due to the continuous nature of the transition in this region and the 
way we define the crossover.

In Case III, we impose that the decay $\sigma\rightarrow\pi\pi$ occurs in the 
isentropic trajectories (Fig. \ref{fig:4}, right panel). 
The location of the CEP is not known yet, but, it can be argued in favor 
of its experimental detection that if the evolution of strongly interacting 
matter is such that the system passes in the vicinity of the CEP starting 
from the initial conditions we will be able to locate it. 
Due to the successes of ideal fluid hydrodynamics at RHIC, it is likely that the 
system expands nearly isentropically and the $\sigma\rightarrow\pi\pi$ decay, 
if it occurs, will be under such conditions.\\
$-$ For $s/\rho_B=9$ (dashed lines) $g_{\sigma\pi\pi}$ and 
$\Gamma_{\sigma\rightarrow\pi\pi}$ drop to zero  and there is a range, 
$176<T<184$ MeV, where the $\sigma\rightarrow\pi\pi$ decay occurs inside the 
first order region and we do not have decays for temperatures and chemical 
potentials above the first order line.\\
$-$ When $s/\rho_B=11$ (dotted lines) $g_{\sigma\pi\pi}$ and 
$\Gamma_{\sigma\rightarrow\pi\pi}$ will be different from zero slightly above the
isentropic line reaches the crossover transition as $T$ decreases. 
The $\sigma\rightarrow\pi\pi$ decay then occurs until $T\approx14$ MeV and for 
$T<14$ MeV it will be inside the first order region.\\
$-$ For $s/\rho_B=10$ (solid red line) the trajectory passes very close to
the CEP. $g_{\sigma\pi\pi}=1.67$ GeV and $\Gamma_{\sigma\rightarrow\pi\pi}=135$ 
MeV in the point near the CEP. 
Both quantities go to zero inside the region of chiral restored phase but near 
the transition line.

When the $\sigma\rightarrow\pi\pi$ decay occurs at the transition lines,
for these values of $s/\rho_B$, 
$g_{\sigma\pi\pi}$ and $\Gamma_{\sigma\rightarrow\pi\pi}$ are smaller when it 
occurs near the CEP: the red dot (at the CEP) in Fig. \ref{fig:4}, right panel, 
has a smaller value compared with the magenta dot (crossover) and the 
blue dot (first order).

The $\sigma\rightarrow\pi\pi$ decay depends on the initial conditions and the 
way the system evolves after the thermalization. 
If the isentropic trajectories reach the first order line for 
$\mu_q>\mu_q^{1st}$, there will be no pions coming from the $\sigma$ 
decay in the chirally symmetric phase. Near the CEP and above the 
$\sigma\rightarrow\pi\pi$ decay is possible and its width is still high near 
the CEP.

\section{Summary}
\label{sec:Summary}

The location of the CEP is one important issue addressed by the HIC program. 
The Nuclotron-based Ion Collider Facility at JINR will significantly enhance 
our understanding of the QCD phase diagram namely the nature of the phase 
transition and the existence/location of the CEP.
The eventual confirmation of the CEP would be one of the first discoveries of 
QCD-like observables in the medium. 
The implications of its location are vast, in particular concerning the 
constraints to set on effective models.

In this work we started to study the isentropic lines in the vicinity of the CEP.
The isentropic trajectories, in the crossover region very close the CEP, show 
``focusing'' effects which can be seen as the result of the CEP to act as an
attractant of isentropic trajectories \cite{Nonaka:2004pg}).  At high chemical
potentials and low temperatures these lines will go through the first order line.

By taking into account the $\pi$ and $\sigma$-mesons properties around the CEP, 
we looked for signatures that can be observed experimentally \cite{Anticic:2009pe}.
We have focused on the $\sigma\rightarrow\pi\pi$ decay that can be suppressed at 
the transition region due to the small mass the $\sigma$-meson can have in that 
region.
This decay showed different behaviors concerning the region where the collision 
takes place and may be used to distinguish different transition types and to 
locate the CEP: if the isentropic trajectories reach the first order region coming 
from higher temperatures, only after the transition the $\sigma\rightarrow\pi\pi$ 
decay happens and no pions from $\sigma$ decay occur in the chirally 
symmetric phase. 
Otherwise, if the isentropic trajectories reach the crossover, 
pions coming from $\sigma$ decay may still occur in the chirally symmetric phase.

\vspace{0.25cm}
{\bf Acknowledgment:} I would like to thank J. Moreira and C. Providência
for helpful discussions.
This work was supported by ``Fundação para a Ciência e Tecnologia", Portugal, 
under the Grant No. SFRH/BPD/1022\ 73/2014.

%
\vspace{-0.25cm}

\end{document}